\documentclass[10pt]{iopart}

\usepackage[english]{babel}
\usepackage{graphicx}
\usepackage{iopams,cite}  
\usepackage{graphicx}

\newcommand{\pdervi}[2]{\partial #1\slash\partial #2}
\newcommand{\derv}[2]{\frac{{\rm d} #1}{{\rm d} #2}}
\newcommand{\dervi}[2]{{\rm d} #1/{\rm d} #2}
\newcommand{\rem}[1]{}

\def\rmd{\mathrm{d}}

\def\oo{\omega}
\def\a{\alpha}

\begin{document}
\title[Collective Thomson scattering diagnostic for the GDT experiment]{Collective Thomson scattering diagnostic for the GDT experiment}
\author{A G Shalashov$^1$, E D Gospodchikov$^1$, T A Khusainov$^1$, L~V~Lubyako$^1$, O B Smolyakova$^1$, A L Solomakhin$^{1,2}$}
\address{$^1$ Institute of Applied Physics of the Russian Academy of Sciences, 46 Ulyanov str., 603950 Nizhny Novgorod, Russia}
\address{$^2$ Budker Institute of Nuclear Physics, 11 Lavrentieva ave., 630090 Novosibirsk, Russia}
\ead{ags@ipfran.ru}

\date{\today}

\begin{abstract}
In this paper, we propose a collective Thomson scattering diagnostic for fast ion measurements for the gas-dynamic trap (GDT) facility at the Budker Institute. The diagnostic utilizes 54.5 GHz gyrotron usually used for electron cyclotron resonance heating as a source of probe radiation and is aimed at reconstruction of distributions over transverse and longitudinal velocities of NBI-driven ions in the plasma core. Here we present a feasibility study of this concept showing a possibility to receive a strong CTS signal of hundreds eVs for a wide range of GDT parameters. The main limitations come from the refraction of the probe and scattered radiation  propagating in inhomogeneous plasma: to provide well-resolved CTS conditions the on-axis plasma density should be kept less than $1.5\cdot10^{13}$\,cm$^{-3}$ while usual GDT discharges correspond to the density of $(0.9-1.3)\cdot10^{13}$\,cm$^{-3}$.  
\end{abstract}


\ioptwocol
\maketitle

\section{Introduction}

The use of microwave radiation for probing the ion velocity distribution with good spatial and temporal resolution has proven its capability for hot plasma confined in toroidal magnetic traps \cite{b1}. Information about ion dynamics is recovered from the scattering of electromagnetic waves on the collective fluctuations of plasma density and, potentially, magnetic field;  such process is shortly referred as a collective Thomson scattering (CTS). Along with optical methods, such as spectroscopy of neutrons and gamma-quanta, the millimeter-wave CTS is one of the main ways of diagnosing the distribution function of fast ions in tokamaks \cite{b14}. The possibilities of CTS were demonstrated experimentally at TFTR \cite{b2}, JET \cite{b3}, TEXTOR \cite{b4,b7}, ASDEX-U \cite{b6aa, b8} tokamaks for diagnosing the distribution function of fast ions, at  W7-AS stellarator for diagnosing the temperature of thermal ions and the lower hybrid plasma instability \cite{b9,b10,b11}, at  LHD stellarator for measurement of both fast and thermal ions \cite{b12,b12aa}, and, most recently, at the newest W7-X stellarator for diagnosing the temperature of thermal ions \cite{b13,b13a}. CTS is considered as a main method for detecting the fusion alpha particles in ITER \cite{b15a,b15}. Another fruitfully developing direction is active microwave diagnostics of small-scale plasma turbulence by measuring the scattering spectra on turbulent density fluctuations \cite{b17,b18,b19,b20}. 

Based on this experience, it seems to be very attractive to exploit the same technique for measuring ion distributions in large open magnetic traps used in magnetic fusion research. Some of devices are already equipped with high-power ECRH systems that may be used a source of probe radiation for the CTS diagnostics \cite{b62,gdt2}. The bulk plasma parameters in the most advanced traps are comparable to those of toroidal machines \cite{prl}, while the population of fast ions and their influence on the performance are usually much greater in open traps than in tokamaks and stellarators. However, microwave CTS diagnostic has not been realized for open traps. A close ideology based on CTS with a CO$_2$-laser as a probing source was planned to be installed at the GAMMA-10 tandem trap for measuring the ion temperature as part of testing the diagnostic system developed for the LHD stellarator \cite{b16}; however, as far as authors know, these experiments did not receive further development.

In this paper we report on the project of the CTS diagnostic for the running experiment at the gas-dynamic trap (GDT) facility at the Budker Institute.

\section{Why we need CTS diagnostic at GDT}

\begin{figure}[b]
\centering 
\includegraphics[width=82mm]{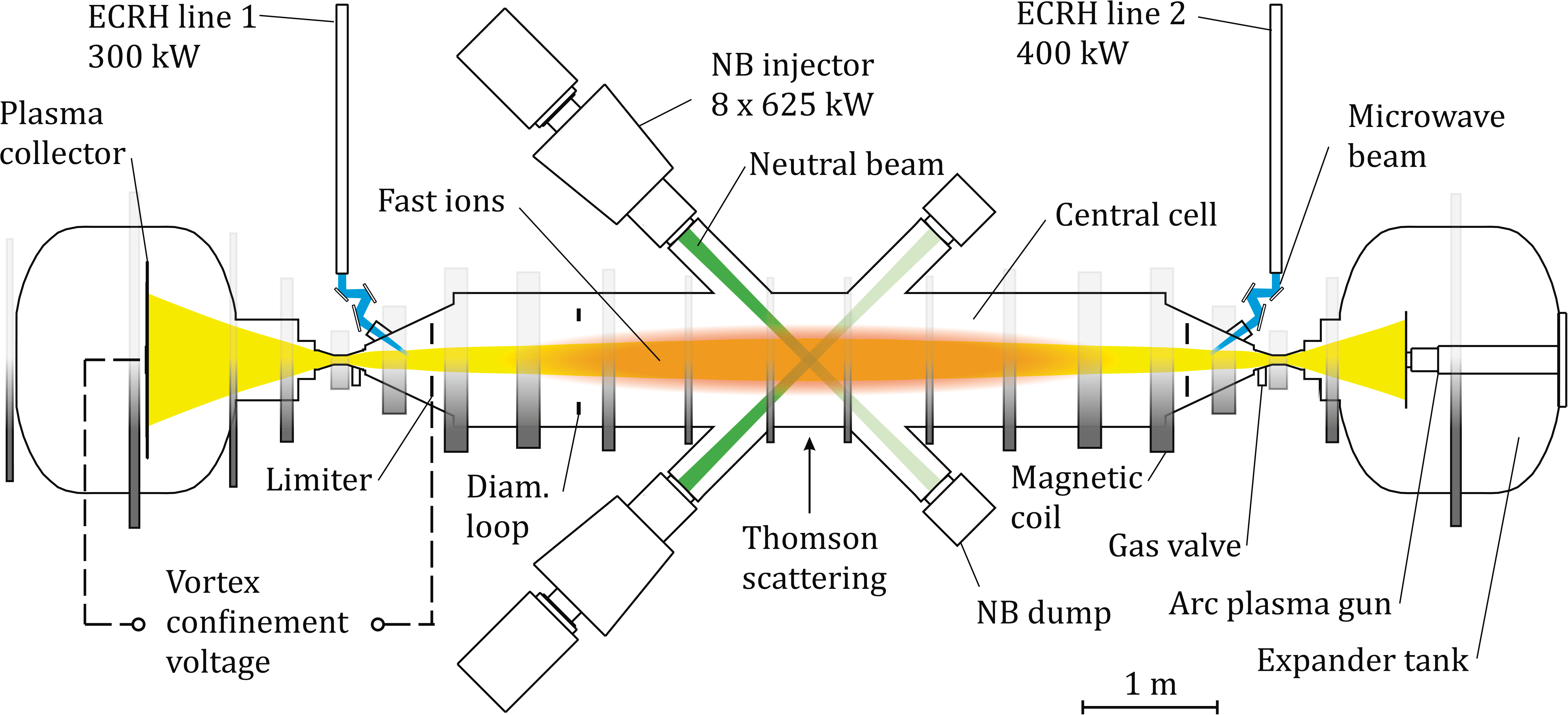}
\caption{ Schematic of the GDT.}\label{fig1} 
\end{figure}

GDT is a fully axisymmetric  linear magnetic device aimed at nuclear fusion applications \cite{iv2013}. The main part of GDT is 7\,m-long central solenoid which is limited by high-field magnetic mirror coils; plasma absorbing end-plates are placed sufficiently far from the magnetic mirrors in a region with expanded magnetic field lines, see figure \ref{fig1}. The plasma heating system consists of eight neutral beam injectors (NBI) providing up to 5 MW of a total injected power and two 400\;kW/54.5\;GHz gyrotrons for the electron cyclotron resonance heating (ECRH). Combined ECRH and NBI heating allows reaching record plasma parameters for large open traps \cite{prl, b12a,gdt1,gdt2,gdt3}. Demonstrated significant progress in the confinement time of fast ions and in the neutron yield (up to 80\%) due to the selective deposition of ECRH power into the electron component of the plasma have led to a noticeable revision of the prospects for using axially symmetric traps as a high-power source of fusion neutrons \cite{sim}. In combination with a number of new theoretical ideas, progress in the field of confinement of hot plasma in a gas-dynamic trap led to a discussion of the possibilities of a fusion power reactor based on the open trap concept \cite{fu0,fu1,fu2}.

In the standard regime of operation, GDT plasma consists of two components. First one is the bulk plasma serving as a target for NBI. This component is confined in the gas-dynamic regime and has an isotropic equilibrium velocity distribution due to a high collision frequency. 
The second plasma component consists of fast ions produced as a result of oblique injection of hydrogen or deuterium atomic beams into the bulk plasma. The fast ions are confined in the adiabatic regime which means that their movement is governed by conservation of energy and magnetic moment (an adiabatic invariant). As a result, they are bouncing in a region between two turning points defined by the effective mirror ratio $R = 2$. The energy confinement time of fast ions is determined by the electron-ion collisions, namely, by the electron drag force; this time turns out to be much less than the angular scattering time. Due to this fact, the fast ions have a strongly non-Maxwellian anisotropic velocity distribution with a relatively small angular spread. Since the distribution function of fast ions is anisotropic in pitch angles, their density is strongly (up to 3 times) peaked near the turning points, what provides favorable conditions for fusion D-D and D-T reactions in subcritical plasma. Thus, the fast ion distribution over pitch angles becomes the main factor determining fusion efficiency, e.g., the neutron yield and the locality of a neutron source  \cite{an00}.   

The ECRH suppresses the main channel of the fast ion loss (collisions with thermal electrons) and changes the anisotropy of the distribution function of fast ions. In this case, the problem of measuring the energy and pitch angle distribution function of fast ions becomes one of the primary ones. In particular, a direct measurement of the distribution function is necessary to clarify existing ideas about the adiabatic nature of fast ion confinement, about the influence on their distribution function of MHD instabilities, electromagnetic instabilities in the ion-cyclotron range, a radial electric field and Coulomb collisions in the new (just achieved for open traps) range of plasma parameters. The strategy of optimizing the GDT operation depends on solving these issues.

The CTS diagnostic with one of the ECRH gyrotrons used as a source of probe radiation seems to be a suitable tool to fulfill this demand. 

Previously, the distribution of fast ions in the GDT plasma was investigated with a charge-exchange (CX) of a diagnostic neutral hydrogen beam \cite{an00,b24}. The 12-channel CX analyzer provided energy resolution of 0.4--1.3 keV and narrow angular resolution below 1$^\circ$ in the angular range of 32$^\circ$-47$^\circ$. The angle in such a system corresponds to the direction of detected ion velocity. Using CX diagnostics, it was possible to measure the time dependence of the distribution function of fast ions in the range of 3--18 keV. However, at each shot, CX energy spectra provide information on the ion distribution in a narrow range of pitch angles.
Due to technical complexity, this method can not be used as a regular diagnosis. 


\section{CTS geometry}

CTS is scattering of electromagnetic waves on plasma density fluctuations. A principle of CTS is sketched in figure 2. The measurements are done using a probe beam with the wave vector $\mathbf{k}^i$ and angular frequency $\oo^i$, which is scattered in the plasma on fluctuations with $(\mathbf{k},\oo)$, and a receiving beam with the wave vector $\mathbf{k}^s$ and frequency $\oo^s$. Three-wave synchronism implies 
\begin{equation}\mathbf{k}=\mathbf{k}^s-\mathbf{k}^i,\quad \oo=\oo^s-\oo^i.\end{equation} Salpeter showed  that if ${k} \lambda_D < 1$, where $\lambda_D$ is the Debye length, the spectrum of the scattering radiation inherits signatures of collective plasma effects \cite{sal}.

\begin{figure}[tb]
\centering 
\includegraphics[width=70mm]{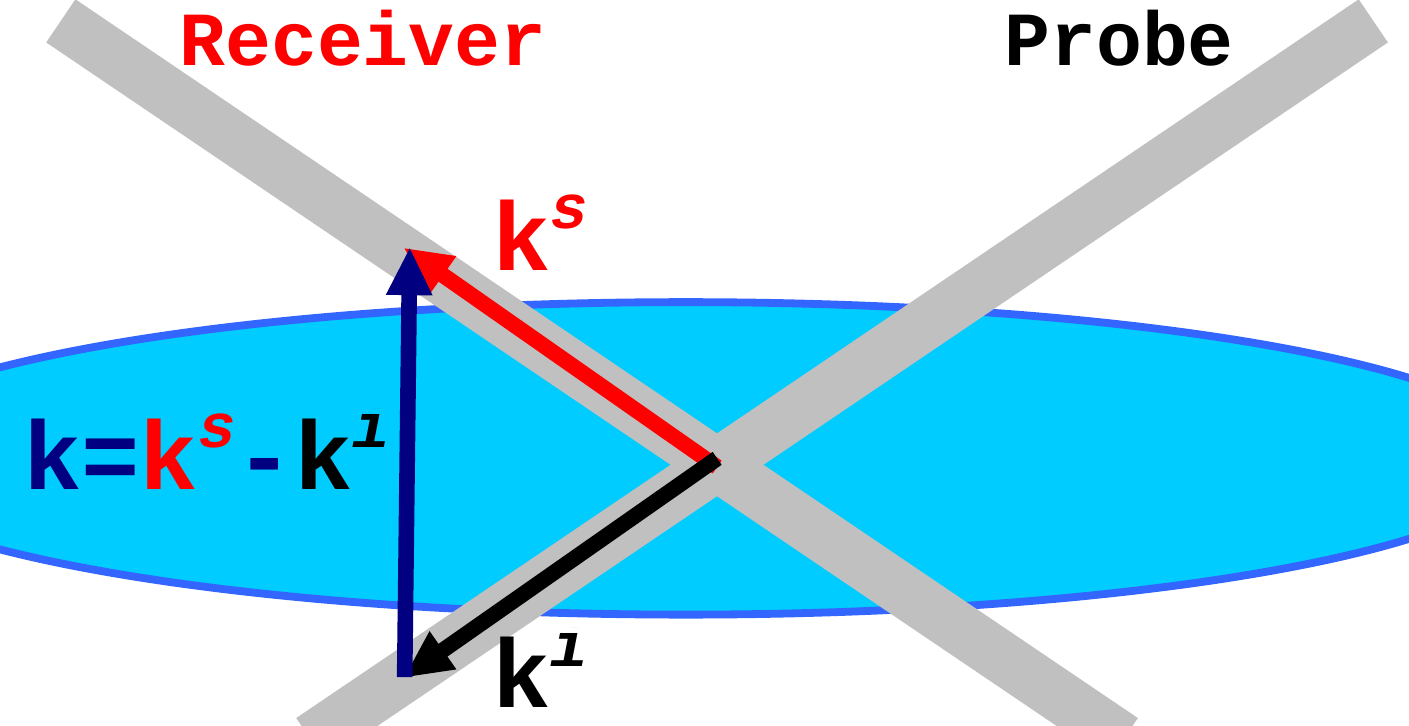}
\caption{ A sketch illustrating a principle of CTS: $\mathbf{k}^i$ and $\mathbf{k}^s$ are the wave vectors of the incident and the received scattered waves, respectively. The fluctuations are measured along $\mathbf{k}=\mathbf{k}^s-\mathbf{k}^i$. The detected fluctuations are collective if ${k} \lambda_D < 1$.}
\label{fig2} \end{figure}

The CTS diagnostic measures a spectrum of scattered microwaves. Let $P^s$ and $P^i$ be the scattered and incident power. The received spectral power density is then given by \cite{a18}
\begin{equation}\label{eq1}
\derv{P^s}{\oo}=P^i \,{\lambda^i\lambda^s}r_e^2 n_e \,O_b\, G\,\frac{S(\mathbf{k},\oo)}{2\pi},	
\end{equation}
where $r_e$ is the classical electron radius, $\lambda^{i,s}=2\pi c/\oo^{i,s}$ are the incident and scattered wavelengths,   $n_e$ is the electron density, $O_b=\int I^i I^s\,\rmd V$ is the beam overlap defined as the volume integral of the normalized incident and scattered beam intensities, $G$ is the geometrical form factor defined by polarization of the incident and scattered radiation, and $S(\mathbf{k},\oo)$ is the spectral density of plasma fluctuations\cite{a19,shef}. 
In the next sections, we will discuss each of these factors, $O_b$, $G$ and $S$, in more detail.

The spectral density of fluctuations consists of two additive parts, $S=S_{t}+S_{f}$, where $S_{t}$ and $S_{f}$ are, respectively, contributions of a thermal bulk plasma and fast ions. In GDT, fast ions can be treated as non-magnetized when calculating the CTS spectra. Then its contribution is proportional to one-dimensional distribution function along the direction of resolved plasma fluctuations: 
\begin{equation}\label{eq2}
S_f\propto F(\oo/k),\quad F(u)=\int f(\mathbf{v})\,\delta(u-\mathbf{k}\mathbf{v}/k)\,\rmd^3\mathbf{v},\end{equation}
where $f(\mathbf{v})$ is a local distribution of fast ions in a three-dimensional velocity space calculated inside the scattering volume. The energy of fast ions is much higher than those of the bulk particles, so fast ions would contribute to the scattering at higher frequencies than thermal ions. In the next sections we will show that there is a frequency domain with $S_f\gg S_t$ favorable for diagnosing the fast ion distribution function. Although recovering of full distribution function $f(\mathbf{v})$ by combining  $F(u)$ along several different lines of sight is, in principle, possible \cite{tomo1,tomo2}, we do not consider such an option in the current GDT project. Thus, the one-dimensional distribution $F(u)$ is an ultimate result of the CTS diagnostic considered in the present paper. 

\begin{table}[b]
\caption{ \label{tab1} Characteristics of GDT CTS (preliminary design).  }
\begin{indented}
\item[]\begin{tabular}{@{}lcc}
\br
	 & CTS I (top) & CTS II (bot.) \\ \mr
Scattering angle $\angle (\mathbf{k}^s,\mathbf{k}^i)$ & 65$^\circ$--87$^\circ$ & 93$^\circ$--110$^\circ$ \\ 
Angle $\angle (\mathbf{k},\mathbf{B})$, $\mathbf{k}\!=\!\mathbf{k}^s\!-\!\mathbf{k}^i$  & 87$^\circ$--93$^\circ$ & 7$^\circ$--16$^\circ$ \\ 
Parallel wavevector $k_{||}$ & $-0.5$--0.5 cm$^{-1}$ & 12--15 cm$^{-1}$\\
Transverse wavevector $k_\bot$ & 11--14 cm$^{-1}$ & 2--3.5 cm$^{-1}$\\
Scattering volume V$_\mathrm{CTS}$	 &900 cm$^3$ & 900 cm$^3$\\
Radial position of V$_\mathrm{CTS}$	 & 5--10 cm & 0--5 cm\\
Axial position of V$_\mathrm{CTS}$	 & $-5$--5 cm & $-5$--5 cm\\
Radial size of V$_\mathrm{CTS}$	 & 4 cm & 4 cm\\
\br
\end{tabular}
\end{indented}\end{table}

\begin{figure}[tb]
\centering 
\includegraphics[width=82mm]{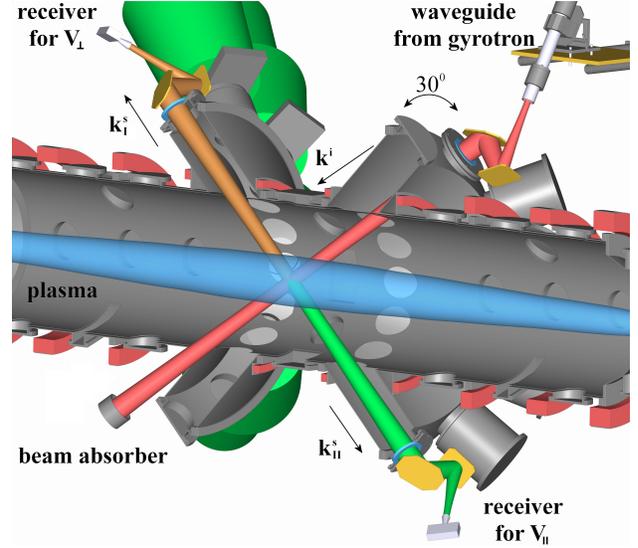}
\caption{ CTS geometry for the GDT setup: $\mathbf{k}^i$ denotes the incident gyrotron beam,  $\mathbf{k}^s_{\mathrm{I}}$ denotes the received CTS beam sensitive to fast ion distribution over perpendicular velocities,  $\mathbf{k}^s_{\mathrm{II}}$ denotes the received CTS beam sensitive to the distribution over longitudinal velocities.}
\label{fig3} 
\end{figure}

Nevertheless, to have a key to actual distributions of fast ions over perpendicular and longitudinal velocities (with respect to an external magnetic field $\mathbf{B}$) we consider two complimentary CTS geometries in which the wave vector $\mathbf{k}=\mathbf{k}^s-\mathbf{k}^i$ is aligned approximately along and across $\mathbf{B}$. Figure \ref{fig3} illustrates the result of adjusting this idea to actual ports available at GDT and of further optimization described in the next sections. The probe gyrotron radiation is launched from the top of a vacuum chamber. The scattered signal is received either from the top or from the bottom. The first case characterized by approximately $(\mathbf{k}^s_{\mathrm{I}}-\mathbf{k}^i)\perp \mathbf{B}$, thus resolves the fast ion distribution over perpendicular velocities. The second case characterized by approximately $(\mathbf{k}^s_{\mathrm{II}}-\mathbf{k}^i) \;||\; \mathbf{B}$, thus resolves the fast ion distribution over longitudinal velocities. Both geometries correspond to the scattering volume in a central part of the trap occupied by the fast ions. To improve coupling, the probe and receiving ports are shifted by $30^\circ$ in azimuthal direction. Characteristics of the proposed scheme are listed in Table \ref{tab1}. The CTS diagnostics implies some limitations on the bulk plasma density. For both CTS schemes, the on-axis density must be less than $1.5\cdot10^{13}$\,cm$^{-3}$ for the reliable operation at the O mode (see below). This is not a critical restriction since the normal on-axis density at GDT lies in the range   of $0.9-1.3\cdot10^{13}$\,cm$^{-3}$.

\section{CTS spectrum}

\begin{figure*}[tb]
\centering 
\includegraphics[width=170mm]{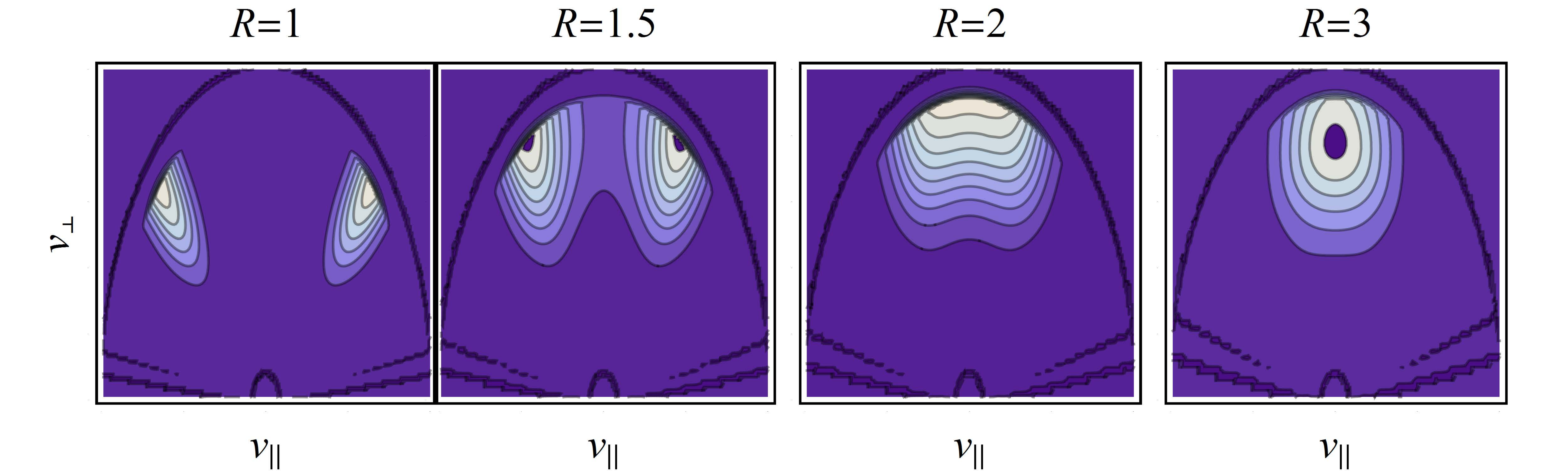}
\caption{  Contour plots of fast ion distribution function $f(v_\bot,v_{||})$ for different positions along the trap axis labeled with the mirror ratios $R=B(z)/B_{\min}$. Maximum velocity in the plot range correspond to 30 keV. Results obtained with DOL code with parameters typical of GDT experiment with pure NBI heating (no ECRH): NBI energy 25 keV, NBI launching angle 45$^\circ$ to trap axis, $B_{\min}=0.35$\;T, $T_e=T_i=180$\;eV, $n_e=10.6\cdot10^{12}\;\mathrm{cm}^{-3}$, $n_i=8\cdot10^{12}\;\mathrm{cm}^{-3}$, $n_f=2.6\cdot10^{12}\;\mathrm{cm}^{-3}$ at $R=1$. Note that these distributions are far from being stationary; calculations stop at time of 4.2 ms just before the real NBI switch-off.}\label{fig4}
\end{figure*}

In GDT conditions, a frequency spectrum of the CTS signal is determined by the  spectral density of fluctuations $S$; all other terms in  \eref{eq1} are slowly varying functions of frequency in a detection band.

For the contribution of bulk plasma we assume isotropic Maxwellian distributions for both electrons and ions. Then the expression for the spectral density function for a magnetized plasma, in the electrostatic approximation, is given by \cite{shef}
\begin{equation}
S_t=\frac{k^2 v_{e}^2}{\oo\oo_{pe}^2}\left\{\left|1-\frac{H_e}{\epsilon_l}\right|^2 \mathrm{Im}\,{H_e}+\frac{T_e}{T_i}\left|\frac{H_e}{\epsilon_l}\right|^2 \mathrm{Im}\,{H_i}\right\},
\end{equation}
where $\epsilon_l$ is the longitudinal dielectric constant,  $H_\a$ denotes  the electron ($\a=e$) and the ion ($\a=i$) susceptibilities, 
\begin{eqnarray}
H_\a=\frac{2\oo_{p\a}^2}{k^2 v_{\a}^2}\times  \\ \nonumber\times \sum_{l=-\infty}^{+\infty}e^{-k_\bot^2\rho_\a^2}I_l(k_\bot^2\rho_\a^2)\left(1+\frac{\oo}{k_{||}v_{\a}}Z\left(\frac{\oo-l\oo_{c\a}}{k_{||}v_\a}\right)\right),
\end{eqnarray}
$\oo_{p\a} =(4\pi e^2n_\a/m_\a)^{1/2}$ and $\oo_{c\a}$ denote the electron and ion plasma and cyclotron frequencies, respectively, $n_\a$ are electron and ion densities, $v_\a=\sqrt{2T_\a/m_\a}$ are the electron and ion thermal velocities, $\rho_\a=v_\a/\oo_{c\a}$ denote the Larmor radii; the parallel and the perpendicular wave vectors are defined with respect to the direction of the magnetic field; $I_l$ are the modified Bessel function of the first kind, and $Z$ is the standard plasma dispersion function. 
Below we always assume pure hydrogen or deuterium plasma, so all ions are singly ionized and $n_e=n_i$. We neglect impurities since (a) little is known about impurities at GDT, and (b) their contribution to CTS spectra is much more narrow than the expected contribution of fast ions which is our final target.   
In GDT conditions, fast ions can be treated as non-magnetized. This essentially simplifies calculation of its contribution to the scattering function and the dielectric constant. Then one obtains
\begin{equation}\label{eqsf}
S_f=\frac{2\pi }{k}\frac{n_f}{n_e}\left|\frac{H_e}{\epsilon_l}\right|^2 F(\oo/k)
\end{equation}
and $\epsilon_l=1+H_e+H_i+G_f$ with 
\begin{equation}
G_f=\frac{4\pi e^2n_f}{m_i k}\int \derv{F}{u}\frac{\rmd u}{\oo-ku+i 0}.
\end{equation}
Here we assume a unit norm for the distribution function, $\int F\rmd u=1$, $n_f$ is a volumetric density of fast ions inside the scattering zone, and $i0$ denotes the Landau bypass rule for $e^{-i\oo t}$ processes.
The term $G_f$, i.e.\ the contribution of fast ions to the plasma shielding, is always small in the frequency range important for the CTS diagnostic (including range where $S_f\gg S_t$). 

\begin{figure}[tb]
\centering 
\includegraphics[width=82mm]{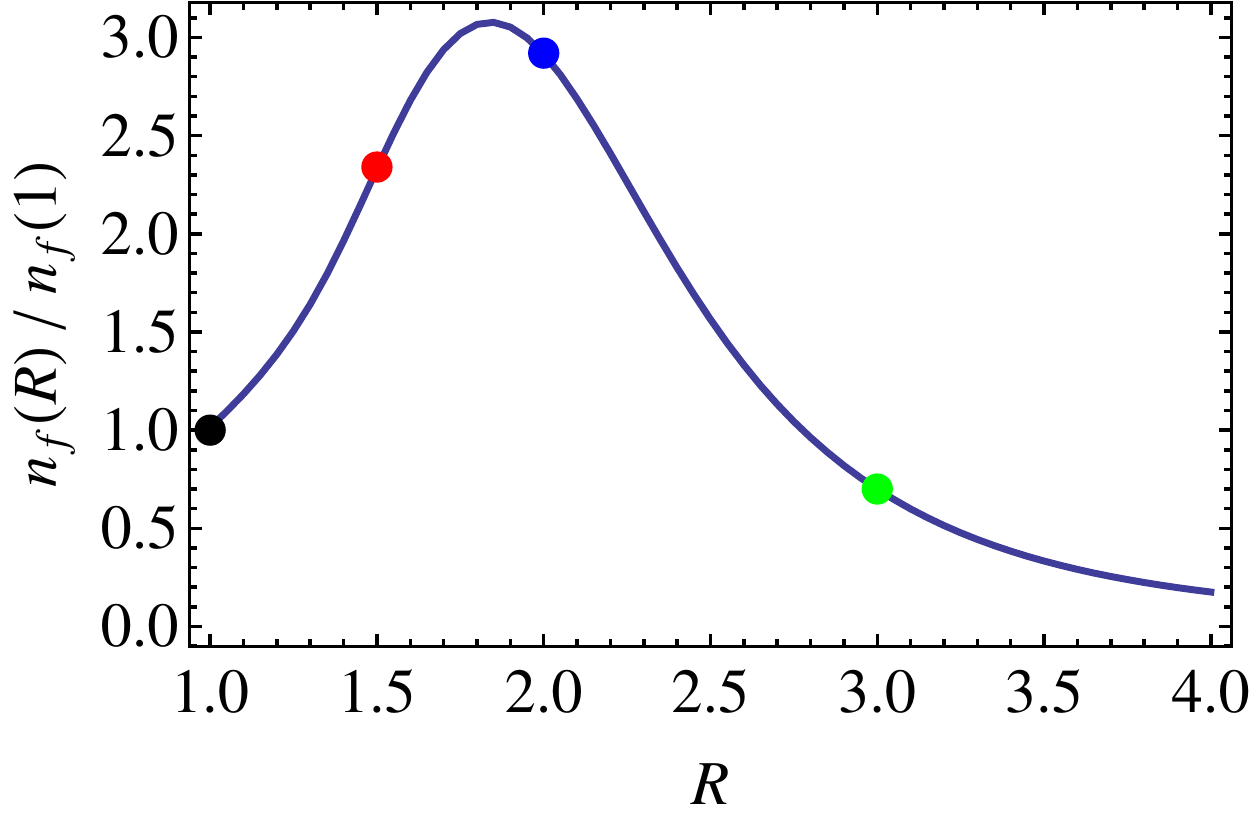}
\caption{ Distribution of fast ion density along the trap axis. Density is normalized over its value in the trap center, plasma conditions are the same as in figure \ref{fig4}. Colored points indicate cases shown in figure \ref{fig7}.} \label{fig5}
\end{figure}

In this paper we perform forward CTS modeling, i.e., we calculate the scattering function for a given distribution function. The fast ion distribution function is calculated with the bounce-averaged Fokker--Planck code DOL, a nonstationary model intended to describe kinetic plasma processes in axisymmetric magnetic mirror traps \cite{DOL}. The output of the code is two-dimensional (axially symmetric in a velocity space) distribution $f(v_\bot,v_{||})$ of collisionally slowed-down NBI-born ions at the magnetic field minimum (corresponded to the trap center). Using invariants of collisionless ion motion along a magnetic field line, $v_\bot^2/B$ and $v_\bot^2+v_{||}^2$, we map the distribution function to other positions along the trap axis. The distribution function is normalized over fast ion density as
\begin{equation}
n_f=\int f(v_\bot,v_{||})\; 2\pi v_\bot \rmd v_\bot\, \rmd v_{||}.
\end{equation}
An example relevant to actual GTD experiments is shown in figure \ref{fig4}. Corresponding density of fast ions along the trap axis is plotted figure \ref{fig5}. One can see that maximum density is reached near the mirror ratio $R=2$ which corresponds to the turning point of an ion born  with the 45$^\circ$ pitch-angle at the trap center. In physical space this point is about 1.8 m far from the center.

\begin{figure*}
\centering 
\includegraphics[width=83mm]{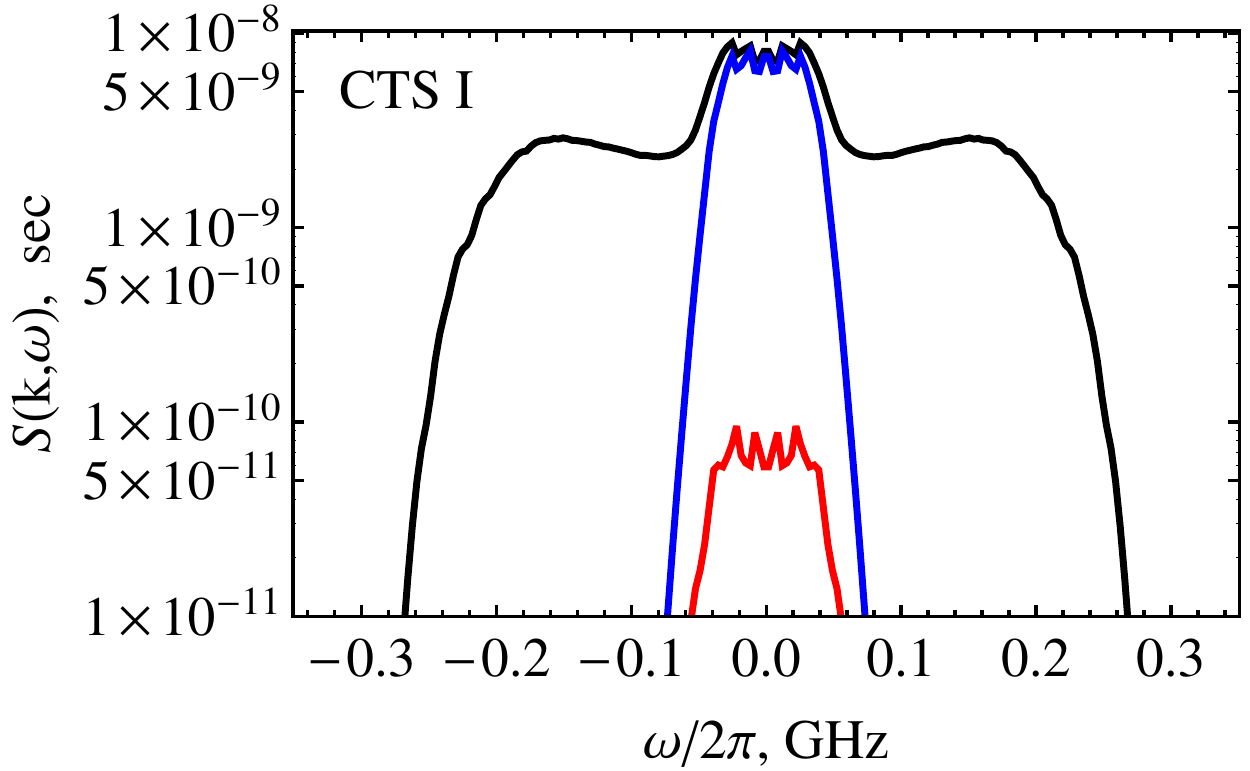}
\includegraphics[width=83mm]{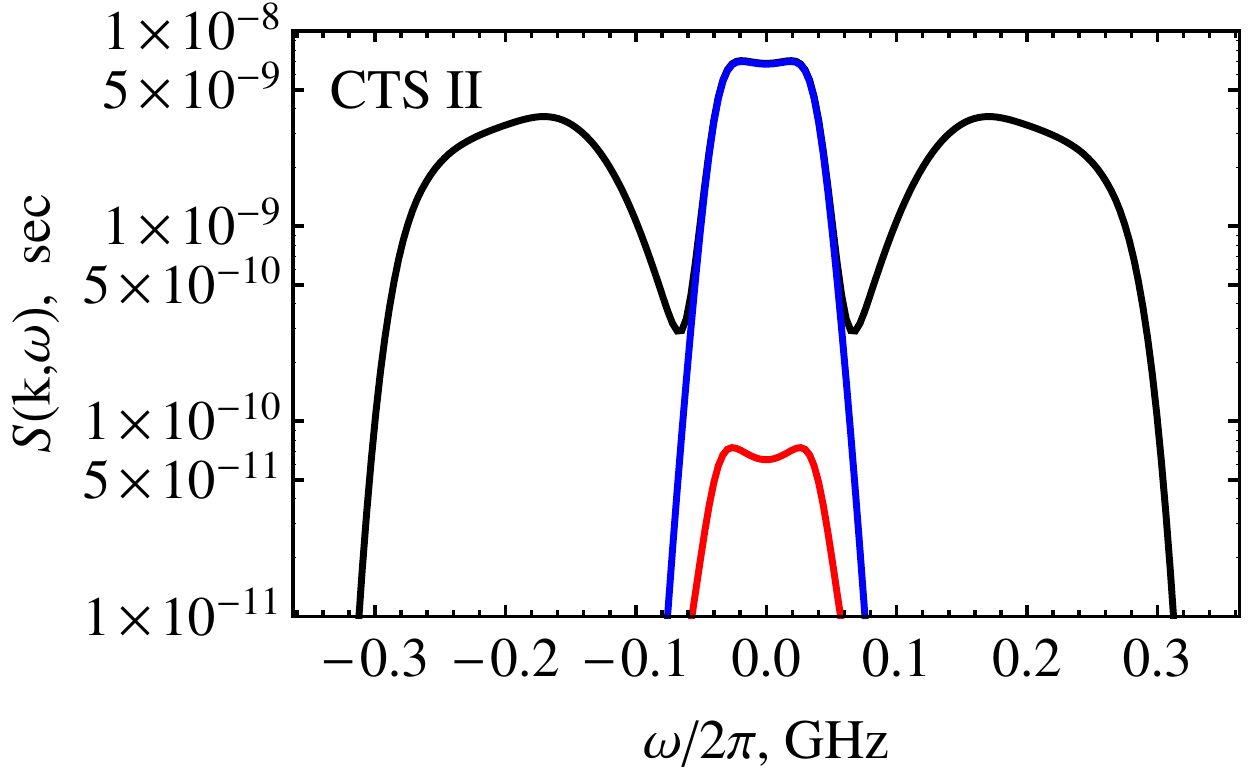}
\caption{ Logarithmic plot of scattering function $S(\mathbf{k},\oo)$ for fast ion distribution at the trap center (black curve). Colored curves show deposition of thermal ions (blue) and electrons (red). Plasma conditions described in figure \ref{fig4}. .} \label{fig6} 
\end{figure*}

\begin{figure*}
\centering 
\includegraphics[width=83mm]{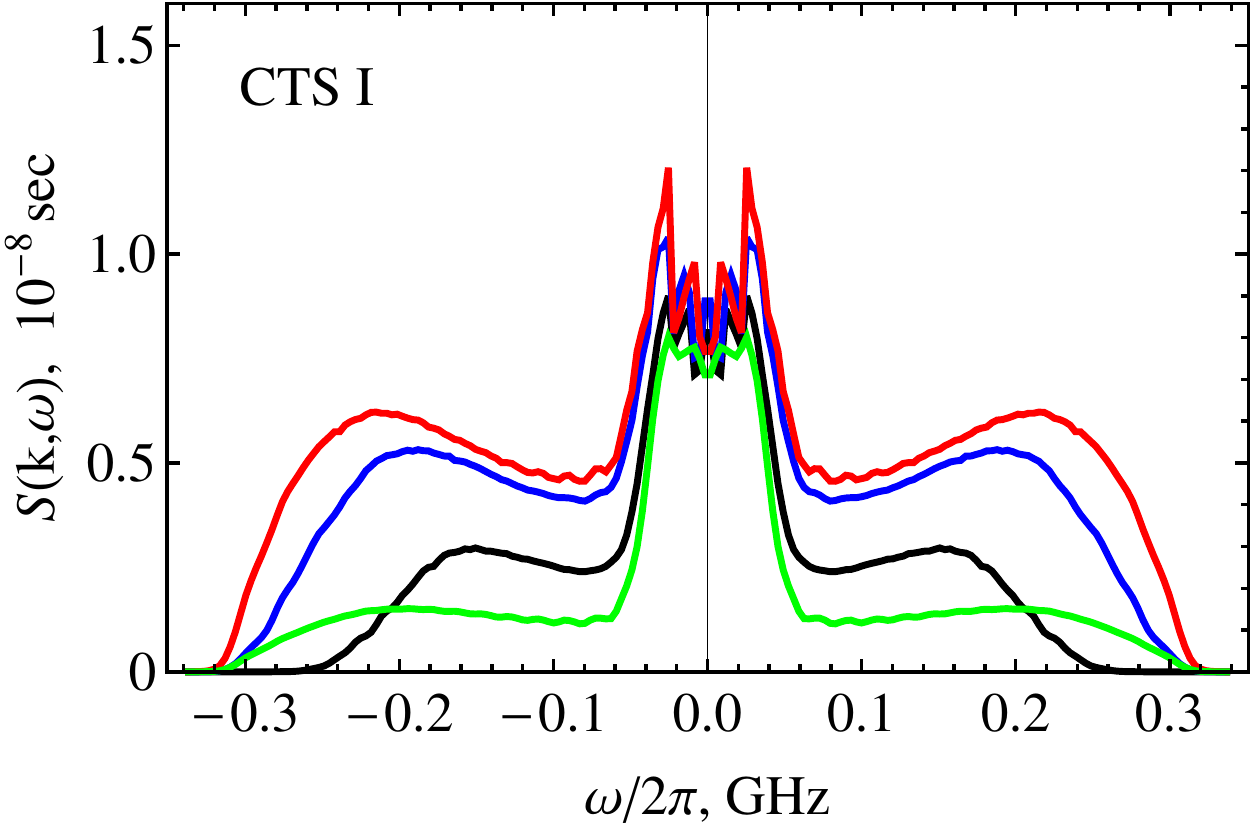}
\includegraphics[width=83mm]{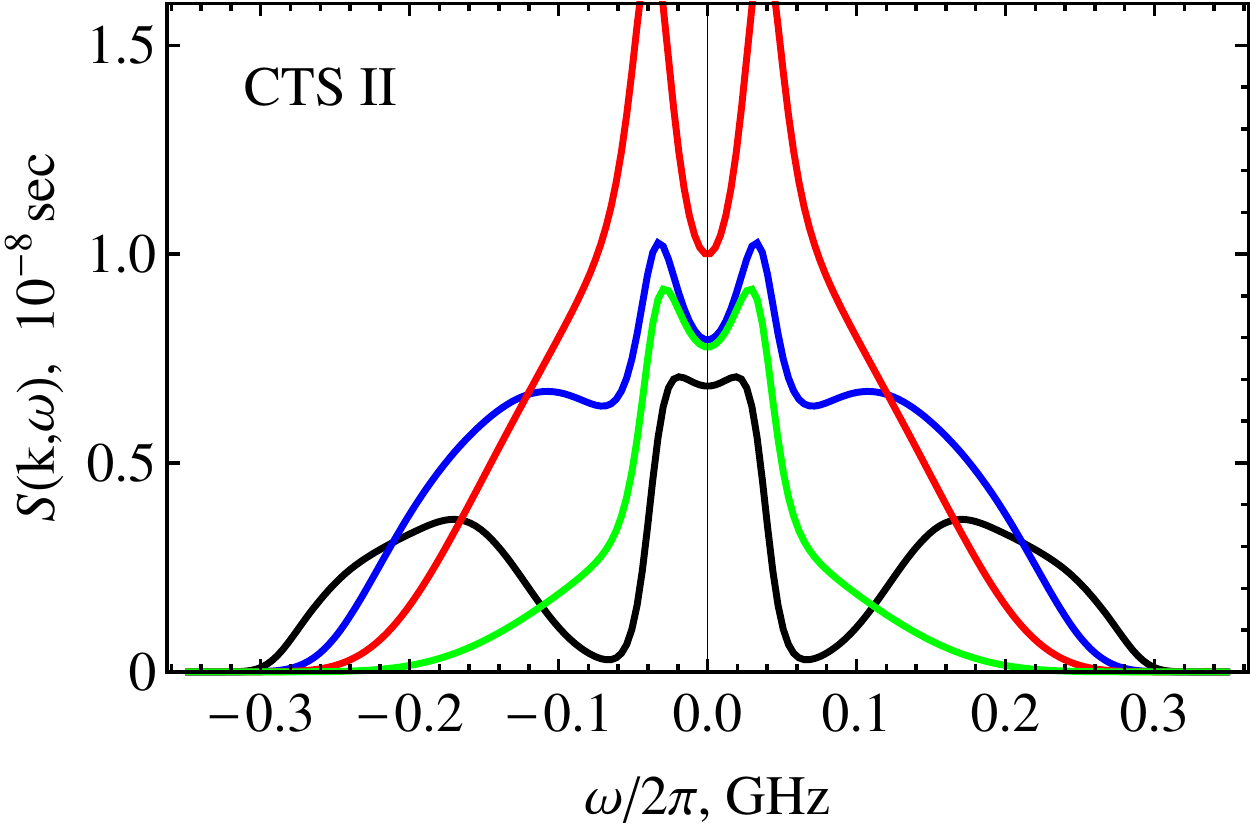}
\caption{ Scattering function $S(\mathbf{k},\oo)$ for fast ion distributions shown in figure \ref{fig4}: black for $R=1$, blue for $R=1.5$, red for $R=2$ and green for $R=3$ (colors correspond to points shown in figure \ref{fig5}).}\label{fig7} 
\end{figure*}

Then we calculate the scattering function $S(\mathbf{k},\oo)$ using  one-dimensional distributions of fast ions obtained as numerical integrals,
\begin{eqnarray}
F(u)=\int f(v_\bot,v_{||})\,\times \nonumber \\ \times\, \delta\left(u-\frac{k_{||}v_{||}+k_\bot v_\bot \cos\phi}{k}\right)\, v_\bot \rmd v_\bot\, \rmd v_{||}\,\rmd \phi.\end{eqnarray}
The results for the CTS geometries I (sensitive to perpendicular velocities) and II (sensitive to longitudinal velocities) are presented in figures \ref{fig6} and \ref{fig7}. Figure \ref{fig6} allows comparing different channels of the scattering. One see that the fast ions are totally dominating in the CTS spectrum in the frequency range of 100--300 MHz. For anisotropic fast ions, the signals of CTS I and II are rather different, while the thermal component results in more or less the same signal in both cases. This is a natural consequence of two facts: the scattering angle is similar in both cases and the magnetic field effect on CTS are weak in our geometry. Some oscillations of CTS I spectrum are due to limitations of our numerical model. 

Figure \ref{fig7} illustrates sensitivity of the CTS spectrum to the shape of the distribution function---different colors correspond to the different positions of the scattering volume along the trap axis (labeled by points with the same colors in figure \ref{fig5}). According to \eref{eqsf}, the CTS signal follows the shape of one-dimensional distribution function $F(\oo/k)$ in the frequency domain $|\oo|\gtrsim 100$ MHz where the contribution of thermal plasma is small.

\section{CTS resolution }

\begin{figure*}
\centering 
\includegraphics[width=120mm]{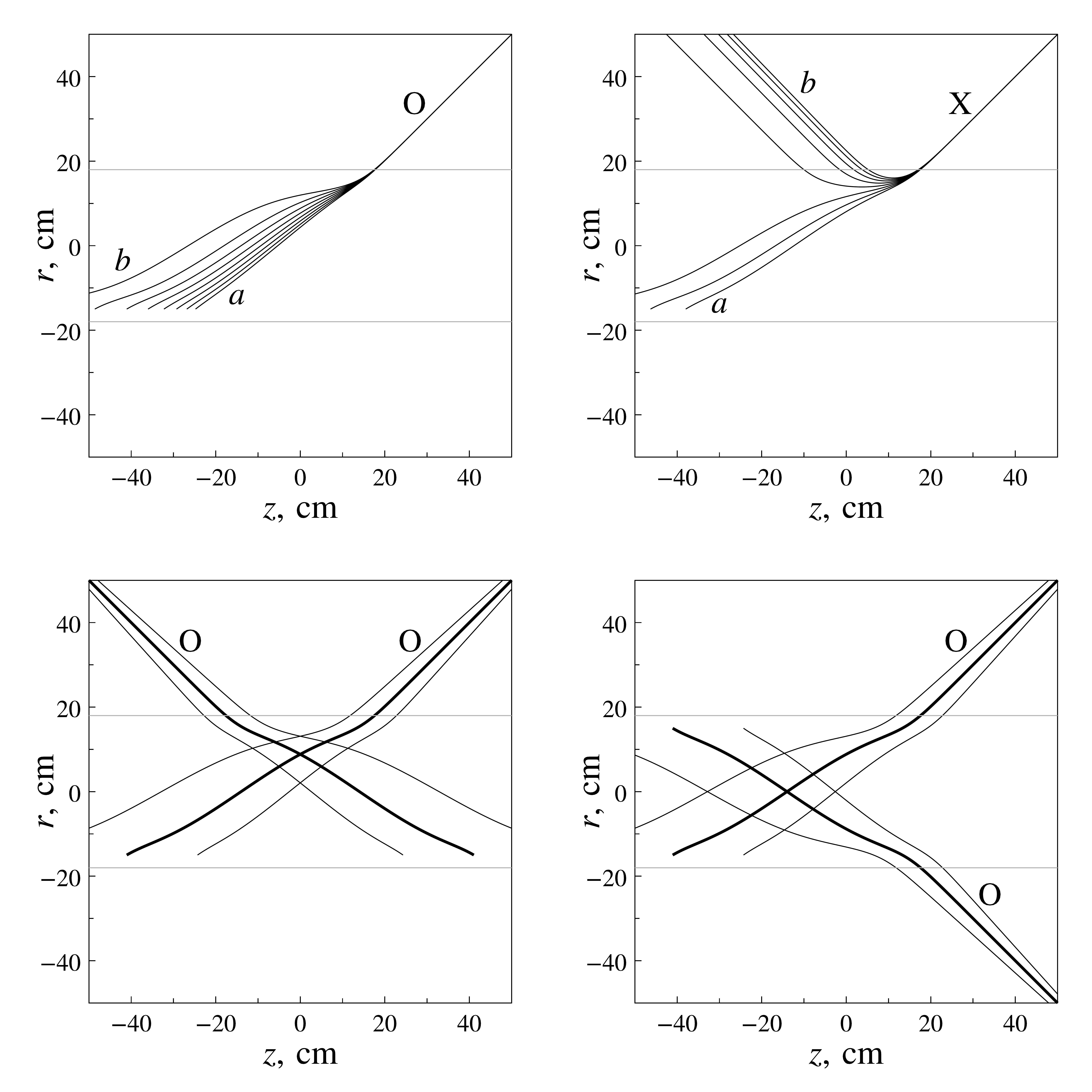}
\caption{ Results of ray-tracing. Top plots: refraction of O and X mode rays as on-axis plasma density varies from ($a$) $0.8\cdot10^{13}\;\mathrm{cm}^{-3}$ to ($b$) $1.5\cdot10^{13}\;\mathrm{cm}^{-3}$. Bottom plots: axial cross-section of CTS I and II scattering geometry for the O-mode. Plasma density and temperature profiles correspond to pure NBI  discharges in GDT and similar to those reported in \cite{pop2017}.}\label{fig8} 
\end{figure*}

In this section we investigate the CTS geometry, namely the beam overlap $O_b$ and the polarization form factor $G$ in \eref{eq1}.

To estimate the influence of radiation refraction in inhomogeneous plasma, we use the ray-tracing code previously developed for the ECRH modeling in GDT \cite{pop2012}. Some results are illustrated in figure \ref{fig8}. The ray-tracing treats independently the O and X modes propagating in the magnetized plasma. Top plots show how the refraction increases with the plasma density: the same ray is launched at a fixed angle for a set of congruent density profiles. One see that plasma cutoffs implies rather strong limitations on operational plasma density: it should be below $\approx10^{13}\;\mathrm{cm}^{-3}$ (on-axis) to avoid the X-mode cutoff and below $\approx1.5\cdot10^{13}\;\mathrm{cm}^{-3}$ to avoid the O-mode cutoff. Based on similar considerations and after checking through all typical GDT conditions, we conclude that O-mode is the only acceptable option for the reliable CTS diagnostic.  

The incident and scattered microwave beams are modeled as a sum over a three-dimensional set of rays with wave intensities, $I^i$ and $I^s$, distributed according to the antenna pattern. The intensity along each ray can be calculated from the trivial radiation transfer equation with no losses, $\dervi{(I|\cos\delta|/N^2)}{l}=0$, where $N$ is a refractive index, $\delta$ is an angle between the group velocity $\pdervi{\oo}{\mathbf{k}}$ and the wave vector,  and $l$ is a coordinate along the ray. Bottom plots in figure \ref{fig8} shows two-dimensional cuts of two crossing three-dimensional beams. Initial conditions for these beams are set with taking into account  restrictions of physical  ports at GDT. In particular, we find that on-axis position of the scattering volume is possible only for the CTS II geometry. By iterating of such calculations for different options available at GDT, we find the restrictions on parameters listed in Table \ref{tab1}. It is interesting to note, that the beam overlapping, calculated as a three-dimensional integral, is practically constant, $O_b=\int I^i I^s\,\rmd V\approx 0.2\;\mathrm{cm}^{-1}$, for all cases of interest. The characteristic size of the scattering volume is about 4 cm what makes 10\% of the actual plasma radius in the central cross-section.

\begin{figure*}
\centering 
\includegraphics[width=56mm]{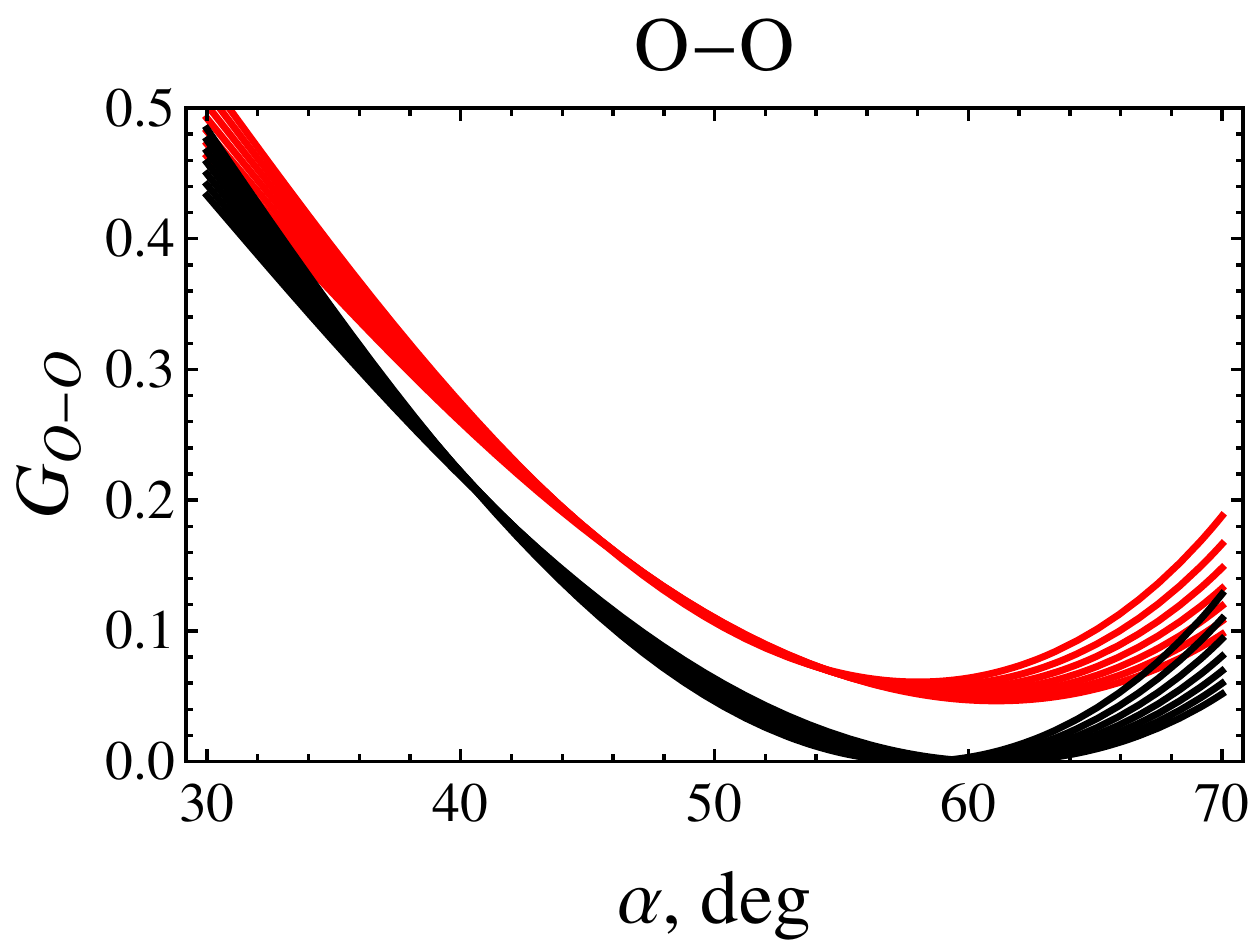}
\includegraphics[width=56mm]{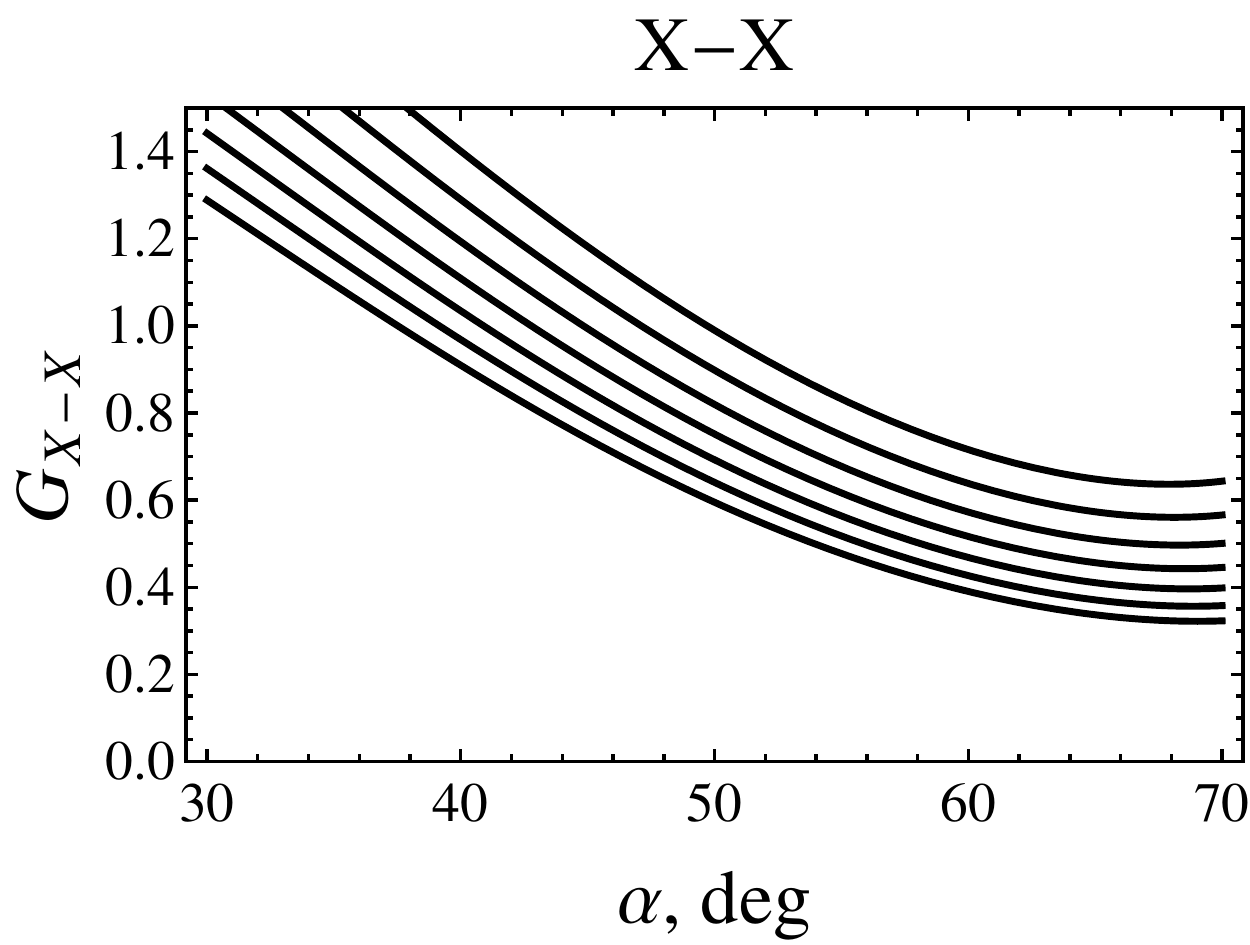}
\includegraphics[width=56mm]{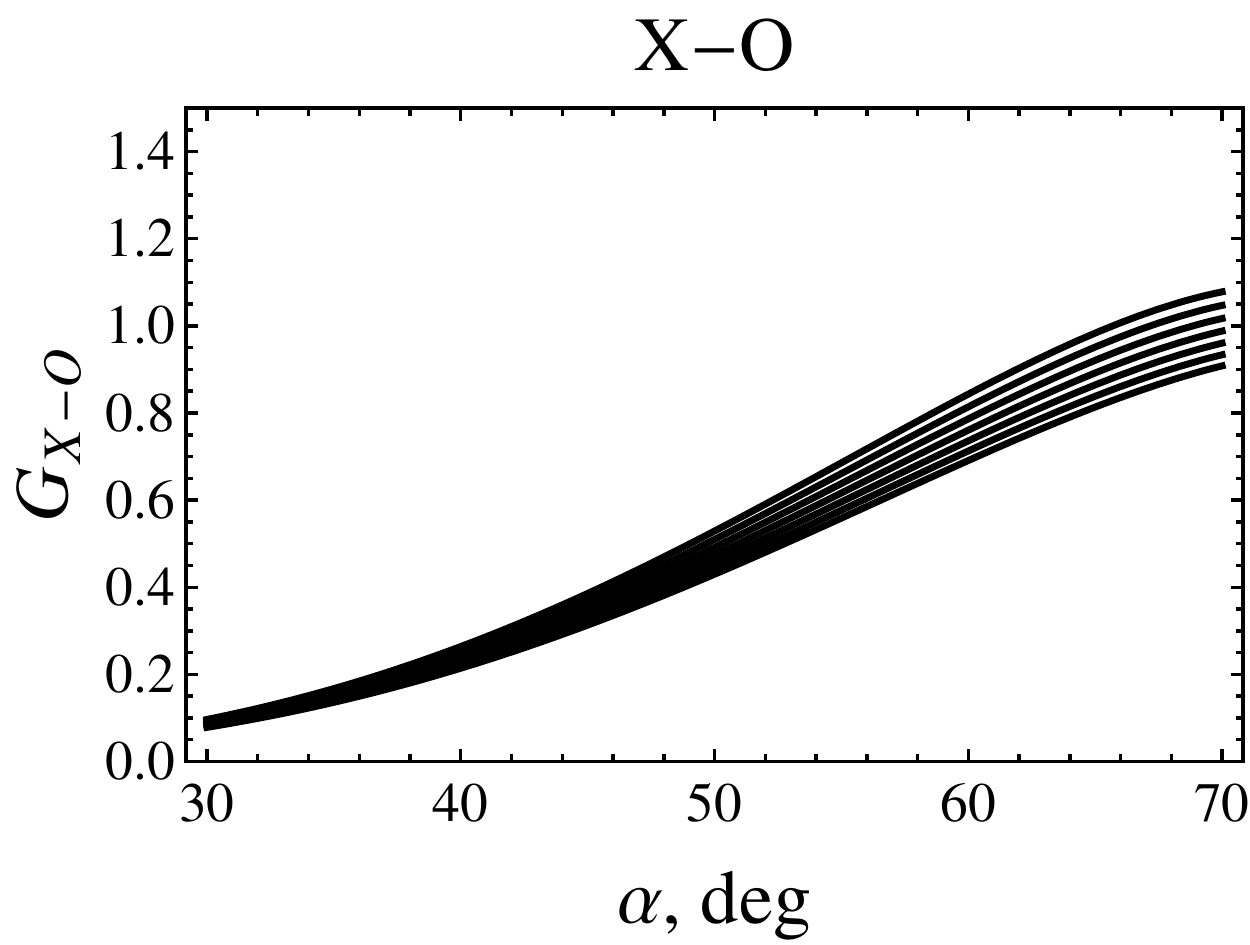}
\caption{Geometrical form factors $G_{\mathrm{O-O}}$, $G_{\mathrm{X-X}}$ and $G_{\mathrm{X-O}}=G_{\mathrm{O-X}}$ as a function of the launching angle $\a$ with respect to the trap axis for pure NBI  discharges in GDT. Set of similar curves corresponds to different on-axis plasma densities that varies from $0.8\cdot10^{13}\;\mathrm{cm}^{-3}$ to $1.4\cdot10^{13}\;\mathrm{cm}^{-3}$. Red curves correspond to the probe and receiving ports shifted by $30^\circ$ in azimuthal direction as shown in figure \ref{fig3}, black curves correspond to the probing and receiving at the same azimuthal position (plain geometry). Physically allowed ranges of $\a$ are $30^\circ-50^\circ$ for the O--O scattering, $20^\circ-45^\circ$ for the X--X scattering, and  $20^\circ-50^\circ$ for the X--O scattering.} \label{fig9} 
\end{figure*}

The geometrical form factor is given by the following expression \cite{a28}:
\begin{eqnarray}
G_{\mu\nu}= \\ \nonumber=\left(\frac{\oo^s\oo^i}{\oo_{pe}^2}\right)^2 \frac{N^s_\nu N^i_\mu|\mathbf{e}^{s*}_\nu\cdot(\hat{I}-\hat{\epsilon}^i)\cdot\mathbf{e}^i_\mu|^2}{(\mathbf{e}^{s*}_\nu\cdot\hat{\epsilon}^s\cdot\mathbf{e}^s_\nu)(\mathbf{e}^{i*}_\mu\cdot\hat{\epsilon}^i\cdot\mathbf{e}^i_\mu)}\, \cos\delta_\nu^s\cos\delta^i_\mu,
\end{eqnarray}
where the subscripts $\nu$ and $\mu$ specify the wave mode (O or X), the superscripts refer to the incident or scattered waves ($i$ or $s$),  $\mathbf{e}$ denotes the polarization vector, $N$ and $\hat{\epsilon}$ are the cold plasma dielectric tensor and corresponding refractive index, respectively.
The results of calculating $G$ are presented in figure \ref{fig9}. Note that using of the O--O scattering would reduce the CTS signal by a factor of 2--5 compared to the most strong X--X scattering; however this is a reasonable pay for the reduced role of refraction when using the O wave polarization. Black curves correspond to a planar scattering geometry when $\mathbf{k}^i$, $\mathbf{k}^i$ and $\mathbf{B}$  lie in the same plane. To improve coupling in the O--O case, one of the ports (probe or receiving) may be rotated in the plane transverse to the trap axis. Red curves in the figure \ref{fig9} show the case when the probe and receiving ports are shifted by $30^\circ$ in azimuthal direction. One can see that this trick, physically possible at GDT, may moderately improve the O--O coupling.

\section{Absolute value of CTS signal}

We analyze an absolute value of the CTS signal as given by \eref{eq1}.  The most variable part in this equation comes from the dependence of the spectral density of plasma  fluctuations $S(\mathbf{k},\oo)$ over $\oo=\oo^s-\oo^i$. This dependence is well illustrated in figures \ref{fig6} and \ref{fig7}. Below we discuss how the vertical axes of these plots scale with other plasma parameters. 

\begin{figure}
\centering 
\includegraphics[width=83mm]{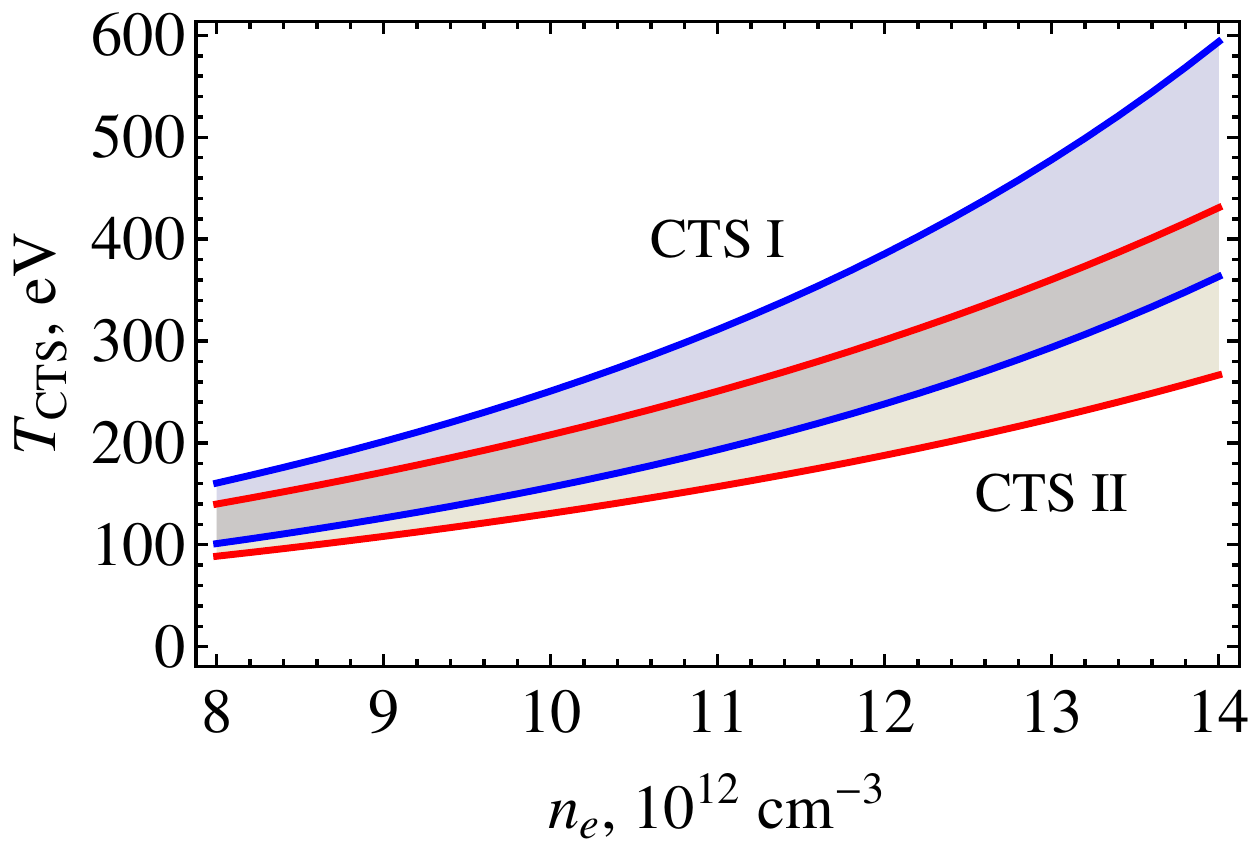}
\caption{ Effective radiation temperature of the CTS signal as a function of plasma density. Finite spread corresponds to variation of the probe and receiving angles in a physically allowed ranges, see Table \ref{tab1}. We assume $S(\mathbf{k},\oo)= 5 \cdot10^{-9}\;$s and the probe power $P^i=400\;$kW at $\oo_i/2\pi=54.5\;$GHz.}\label{fig10} 
\end{figure}

The CTS signal is conventionally characterized in terms of the effective radiation temperature,  \begin{equation}T_{CTS}/ 2\pi= \dervi{P^s}{\oo}.\end{equation} To avoid fast variation of $T_{CTS}$ with $\oo$, we assume a characteristic value $S\approx 5 \cdot10^{-9}\;$s  while calculating \eref{eq1}. All other terms  may be calculated with ray-tracing as discussed in the previous section.  The final result of searching in the allowed parameter range is presented in figure \ref{fig10}. 

The expected level of the CTS signal is of 100--500 eV in 300 MHz frequency band or, equivalently, power of 5--25 nW at the receiver input. These are rather  high values compared to toroidal fusion experiments; we see at least two reasons of such difference. First, as already mentioned in the Introduction, the GDT experiment  is characterized by essentially higher densities of explored fast ions compared to most experiments in toroidal devices. The second reason is lower frequency of the probe radiation.  One finds that $T_{CTS}\propto S/(k^i k^s) $ from \eref{eq1} and the scattering function scales approximately as $S\propto 1/k$, e.g., from \eref{eqsf} and neglecting the variation over difference frequency $\oo=\oo^s-\oo^i$. Then, noting that $k,k^i,k^s\propto \oo^i$, we derive $T_{CTS}\propto (\oo^i)^{-3}$. So, while we consider probing at approximately 55 GHz, modern ECRH systems of tokamaks and stellarators are usually operated at 110--170 GHz (any frequency below requires a dedicated gyrotron, like the one for the ITER project). This gives us a benefit factor from 10 to 30. 

Note that the CTS volume is located far from the electron cyclotron-resonance at the fundamental harmonic, approximately near the 6-th cyclotron harmonic. With the electron temperature of 200--600 eV, the electron-cyclotron emission is virtually negligible, about $10^{-6} \,T_{CTS}$ due to transport of radiation at the fundamental harmonic. 

The main sources of noise in the proposed diagnostic come from the receiver itself and from the gyrotron used as a source of probe radiation. Although we do not discuss here hardware implementation, we mention that the  receiver developed for the CTS diagnostics has noise less than 1 eV. Thus we expect a good signal-to-noise ratio (SNR) in CTS spectra with required frequency resolution about 5 MHz even for a strongly non-stationary GDT plasma with a characteristic time of $\sim$\,1 ms. As in all other CTS experiments, the receiver must be secured from high-power radiation of a probe gyrotron with a dedicated rejection filter (notch-filter). Measurements for the present GDT ECRH system reveal that the gyrotron frequency oscillates during a plasma shot with an amplitude of about 5 MHz: these oscillation caused by the existing power supply and, in principle, may be further reduced. These oscillations give us a minimal spectral width of the notch-filter. To be safe and to provide a possibility of operation with two gyrotrons, one for probing and one for plasma heating, we currently consider a multicavity  notch-filter providing at least 40 dB at $\Delta f<25$ MHz similar to the one described in \cite{lub1}. Another possible source that may effect SNR comes from a gyrotron noise spectrum out of the main generation line but in the CTS band \cite{lub2}; this issue requires additional experimental investigations. 

\section{Summary}

The aim of this work is to identify the conditions under which a collective Thomson scattering experiment with available 54.5 GHz gyrotron source can, in principle, provide useful information about the fast ion particle velocity distribution in GDT. 
We find that it is possible with quite modest hardware requirements when operating in central regions of the trap characterized by low cyclotron plasma emission. The design of components for the CTS diagnostic is now in progress. Compared to well developed experiment in toroidal devices, we have two factors that eventually simplify the diagnostic: much higher relative density of explored fast ions and lower frequency of the probe radiation that increases the scattering efficiency. 
On the other hand, GDT discharge is short-time ($5-10\;$ms) and essentially nonstationary; so there is a limited possibility to collect the CTS signal in order to improve sensitivity, a usual technique in big toroidal machines. 

Although we do not present here a comparison to only competing radiation source based on CO$_2$-laser, a similar research for ITER shows the preference for the gyrotron source \cite{b15}. At 10.6 nm, the CTS spectrum changes very rapidly with angle, only small-angle scattering can be used. In contrast, at gyrotron frequencies, much larger scattering angles are available, the spectra change slowly with angle, and reasonably large collection solid angles granting the Salpeter condition can be used. These benefits are fully exploited in our project. 

\ack The authors thank Peter Bagryansky and the GDT team for hospitality and many stimulating discussions, and Vadim Prikhodko for the help with Fokker--Planck simulations of fast ion distributions. 
The work has been supported by Russian Science Foundation (grant No.~19--72--20139).

\section*{References}
\def\etal{\textit{et al.}}


\begin{thebibliography}{99}


\bibitem{b1}   Bindslev H 1999 \textit{ Rev. Sci. Instrum. }\textbf{70} 1093 

\bibitem{b14}  Moseev D, Salewski M, Garcia-Mu\~noz M, Geiger B and Nocente M 2018 \textit{Rev. Mod. Plasma Phys.}  \textbf{2:7} 

\bibitem{b2}  Machuzak J S, Woskov P P, Gilmore J, Bretz N L, Park H K and Bindslev H 1997 \textit{Rev. Sci. Instrum. }  \textbf{68} 458 

\bibitem{b3}  Bindslev H, Hoekzema J A, Egedal J, Fessey J A, Hughes T P and Machuzak J S 1999 \textit{Phys. Rev. Lett.}  \textbf{83} 3206 

\bibitem{b4}  Porte L, Bindslev H, Hoekzema F, Machuzak J, Woskov P and Van Eester D 2001 \textit{Rev. Sci. Instrum.}  \textbf{72} 1148 



\bibitem{b7}  Stejner M, Nielsen S K, Korsholm S B, Salewski M, Bindslev H, Furtula V, Leipold F, Meo F, Michelsen P K, Moseev D, B\"urger A, Kantor M and de Baar M 2010 \textit{Rev. Sci. Instrum.}  \textbf{81} 10D515 

\bibitem{b6aa} Salewski M, Meo F, Stejner M, Asunta O, Bindslev H, Furtula V, Korsholm S B, Kurki-Suonio T, Leipold F, Leuterer F, Michelsen P K, Moseev D, Nielsen  S K , Stober J, Tardini G, Wagner D, Woskov P. and the ASDEX Upgrade team 2010 \textit{Nucl. Fusion}  \textbf{50} 035012 

\bibitem{b8}  Meo F, Bindslev H, Korsholm S B, Furtula V, Leuterer F, Leipold F, Michelsen P K, Nielsen S K , Salewski M, Stober J, Wagner D and Woskov P 2008  \textit{Rev. Sci. Instrum.}  \textbf{79}  10E501 

\bibitem{b9}  Suvorov E V, Holzhauer E, Kasparek W, Lubyako L V, Burov A B, Dryagin Y A, Fil'chenkov S E, Fraiman A A, Kukin L M, Kostrov A V \etal 1997 \textit{Plasma Phys. Control. Fusion } \textbf{39}  B337 

\bibitem{b10}  Suvorov E V, Holzhauer E, Kasparek W, Burov A B, Dryagin Y A, Fil'chenkov S E, Fraiman A A, Lubyako L V, Ryndyk D A, Skalyga N K \etal 1998 \textit{Nucl. Fusion}  \textbf{38}  661 

\bibitem{b11}  Shalashov A G,  Suvorov E V,  Lubyako L V,  Maassberg  H and W7-AS Team 2003 \textit{Plasma Phys. Control. Fusion} \textbf{45} 395 


\bibitem{b12}  Kubo S, Nishiura M, Tanaka K, Shimozuma T, Yoshimura Y, Igami H, Takahash H, Mutoh T, Tamura N, Tatematsu Y, Saito T, Notake T, Korsholm S B, Meo F, Nielsen S K, Salewski M and Stejner M 2010 \textit{Rev. Sci. Instrum.} \textbf{81} 10D535 
\bibitem{b12aa} Nishiura M, Kubo S, Tanaka K, Seki R, Ogasawara S, Shimozuma T, Okada K, Kobayashi S, Mutoh T, Kawahata K, Watari T, LHD Experiment Group, Saito  T, Tatematsu Y, Korsholm S B and Salewski M 2014 \textit{Nucl. Fusion} \textbf{54} 023006 

\bibitem{b13} 
Moseev D, Stejner M, Stange T, Abramovic I, Laqua H P, Marsen S, Schneider N, Braune H, Hoefel U, Kasparek W, Korsholm S B, Lechte C, Leipold F, Nielsen S K, Salewski M, Rasmussen J, Weissgerber M, Wolf R C 2019 \textit{Rev. Sci. Instrum.} \textbf{90} 013503

\bibitem{b13a} Abramovic I, Pavone A, Moseev D, Lopes Cardozo N J, Salewski M, Laqua H P, Stejner M, Stange T, Marsen S, Nielsen S K, Jensen T, Kasparek W and W7-X Team 2019 \textit{Rev. Sci. Instrum. }\textbf{90} 023501

\bibitem{b15a}  Hughes T and Smith S 1988\textit{  Nucl. Fusion} \textbf{28}(8) 1451 

\bibitem{b15} Bindslev H, Meo F, Tsakadze E L, Korsholm S B and Woskov P 2004 \textit{Rev. Sci. Instrum. A}  \textbf{75} 3598 

\bibitem{b17} Batanov G M, Kolik L V, Sapozhnikov A V, Sarkisyan K A, Skvortsova N. N. and Shats M G  1986 \textit{Fiz. Plazmy} \textbf{12}  1027 [1986 Sov.  J.  Plasma  Phys. \textbf{12} 587]

\bibitem{b18}  Batanov G M, Kolik L V, Petrov A E , Pshenichnikov A A, Sarksyan K A, Skvortsova N N, Kharchev N K, Khol'nov Yu V, Okubo K, Shimozuma T, Ioshimora I, Kubo S, Sanchez J, Estrada T and van Milligen B 2003 \textit{Plasma Phys. Rep.}  \textbf{29}(5)  363 

\bibitem{b19} Kharchev N, Tanaka K, Kubo S, Igami H, Batanov G, Petrov A, Sarksyan K, Skvortsova N, Azuma Yo and Tsuji-Iio S  2008 \textit{Rev. Sci. Instrum.} \textbf{79} 10E721 

\bibitem{b20} Skvortsova N N, Akulina D K, Batanov G M, Kharchev N K, Kolik L V, Kovrizhnykh L M, Letunov A A, Logvinenko V P, Malakhov D V, Petrov A E, Pshenichnikov A A, Sarksyan K A and Voronov G S 2010 \textit{Plasma Phys. Control. Fusion} \textbf{52} 055008 

\bibitem{b62}  Yoshikawa M, Yasuhara R, Nagasu K, Shimamura Y, Shima Y, Kohagura J, Sakamoto M, Nakashima Y, Imai T, Ichimura M, Yamada I, Funaba H, Kawahata K and Minami T 2014\textit{ Rev. Sci. Instrum.} \textbf{85} 11D801

\bibitem{gdt2} Bagryansky P A, Anikeev A V, Denisov G G, Gospodchikov E D, Ivanov A A, Lizunov A A, Kovalenko Y V, Malygin V I, Maximov V V, Korobeinikova O A, Murakhtin S V, Pinzhenin E I, Prikhodko V V, Savkin V Y, Shalashov A G, Smolyakova O B, Soldatkina E I, Solomakhin A L, Yakovlev D V and Zaytsev K V 2015 \textit{Nucl. Fusion} \textbf{55} 053009
\bibitem{prl} Bagryansky P A, Shalashov A G, Gospodchikov E D, Lizunov A A, Maximov V V, Prikhodko V V, Soldatkina E I, Solomakhin A L and Yakovlev D V 2015  \textit{Phys. Rev. Lett.} \textbf{114} 205001
\bibitem{b16}  Kondoh T, Hayashi T, Kawano Y, Kusama Y, Sugie T, Hirata M and Miura Y 2007 \textit{Fusion Sci. Technol.} \textbf{51}(2T) 62 


\bibitem{iv2013}  Ivanov  A A and  Prikhodko V V 2013  \textit{Plasma Phys. Control. Fusion}  \textbf{55} 063001
\bibitem{b12a}  Bagryansky P A,  Kovalenko Yu V,  Savkin V Ya,  Solomakhin A L and  Yakovlev D V 2014  \textit{Nucl. Fusion} \textbf{54}  082001
\bibitem{gdt1} Bagryansky P A, Gospodchikov E D, Kovalenko Y V, Lizunov A A, Maximov V V, Murakhtin S V, Pinzhenin E I, Prikhodko V V, Savkin V Y, Shalashov A G, Soldatkina E I, Solomakhin A L and Yakovlev D V 2015\textit{ Fusion Sci. Technol.} \textbf{68} 87
\bibitem{gdt3}  Yakovlev D V, Shalashov A G, Gospodchikov E D, Maximov V V, Prikhodko V V, Savkin V Ya, Soldatkina E I, Solomakhin A L and Bagryansky P A 2016  \textit{Nucl. Fusion} \textbf{58} 094001
\bibitem{sim} Simonen T C 2016 \textit{J. Fusion Energy} \textbf{35}(1) 
\bibitem{fu0} Gota H, Binderbauer M W, Tajima T, Putvinski S, Tuszewski M, Dettrick S, Garate E, Korepanov S, Smirnov A, Thompson  M C \etal, 2017  \textit{Nucl. Fusion} \textbf{57}  116021 
\bibitem{fu2}   Bagryansky P A,  Beklemishev A D and Postupaev V V 2018 \textit{J. Fusion Energy} \textbf{38} 162
\bibitem{fu1}  Bagryansky P A, Gospodchikov E D, Ivanov A A, Lizunov A A, Kolesnikov E Yu, Konshin Z E, Korobeynikova A A, Kovalenko Yu V, Maximov V V, Murakhtin S V, Pinzhenin E I, Prikhodko V V, Savkin V Ya, Shalashov A G, Skovorodin D I, Soldatkina E I, Solomakhin A L and Yakovlev  D V 2019 \textit{Plasma Fusion Res.}  \textbf{14} 2402030
\bibitem{an00} Anikeev A V, Bagryansky P A, Ivanov A A, Karpushov A N, Korepanov S A, Maximov V V, Murakhtin S V, Smirnov A Yu, Noack K and Otto G 2000 \textit{Nucl. Fusion} \textbf{40} 753
\bibitem{b24} Prikhodko V V , Anikeev A V, Bagryansky P A, Lizunov A A, Maximov V V, Murakhtin S V and Tsidulko Yu A 2005 \textit{Plasma Phys. Rep.} \textbf{31}(11)  899 
 
\bibitem{sal} Salpeter E E 1960 \textit{Physical Review} \textbf{120} 1528
\bibitem{a18} Bindslev H \textit{On the theory of Thomson scattering and reflectometry in a relativistic magnetized plasma}, Ph.D. thesis, Riso National Laboratory, 1992 


\bibitem{shef} Sheffield J \textit{Plasma Scattering of Electromagnetic Radiation} (Academic Press, New York, 1975)
\bibitem{a19} Sheffield J, Froula D, Glenzer S H and Luhmann Jr. N C \textit{Plasma Scattering of Electromagnetic Radiation: Theory and Measurement Techniques}, 2nd ed. (Elsevier, 2010)

\bibitem{tomo1} Salewski M, Nielsen S K, Bindslev H, Furtula V, Gorelenkov N N, Korsholm S B, Leipold F, Meo F, Michelsen P K, Moseev  D and Stejner M 2011 \textit{Nucl. Fusion} \textbf{51} 083014

\bibitem{tomo2} Salewski M, Geiger B, Nielsen S K, Bindslev H, Garcia-Mu\~noz M, Heidbrink W W, Korsholm S B, Leipold F, Meo F, Michelsen P K, Moseev D, Stejner M, Tardini  G and the ASDEX Upgrade team 2012 \textit{Nucl. Fusion} \textbf{52} 103008

\bibitem{DOL}  Yurov D V,  Prikhodko V V and  Tsidulko Yu A 2016 \textit{Plasma Phys. Rep.} \textbf{42} 210
\bibitem{pop2012} Shalashov A G, Gospodchikov E D, Smolyakova O B, Bagryansky P A, Malygin V I, and Thumm M 2012 \textit{Phys. Plasmas} \textbf{19} 052503

\bibitem{pop2017} Shalashov A G, Solomakhin A L , Gospodchikov E D, Lubyako L V, Yakovlev D S, Bagryansky P A 2017 \textit{Phys. Plasmas} \textbf{24} 082506

\bibitem{a28} Hughes T P and Smith S R P 1989 \textit{J. Plasma Phys.} \textbf{42} 215 

\bibitem{lub1} Dryagin Yu, Skalyga N, and Geist T 1996 \textit{Int. J. of Infrared and MM Waves} \textbf{17} (7) 1199

\bibitem{lub2} Suvorov E V, Kasparek W, Lubyako L V, Skalyga N K, Erckmann V, and Laqua H 1998 \textit{Proc. 3-d Int. Kharkov Symposium Physics and Engeneering of Millimeter and Submillimeter Waves} vol 1 (Kharkov: KhPTI) p 188


\end{thebibliography}
\end{document}